\documentclass[smallextended]{svjour3}
  \smartqed
 \usepackage{pdflscape}
 \usepackage[
	pdftex,
	pdfpagemode={UseOutlines},
	bookmarks,
	bookmarksopen,
	colorlinks,
	linkcolor={blue},
	citecolor={green},
	urlcolor={red}
]{hyperref}
 \usepackage[english]{babel}
 \usepackage[T1]{fontenc}
 \usepackage{times}
 \usepackage[sort,authoryear]{natbib}
 \usepackage{mathptmx}
 \usepackage[dvips]{graphicx}
 \usepackage{amssymb}
 \usepackage{amsmath}				%
 \usepackage{multirow}
 \journalname{Experimental Astronomy}
 \graphicspath{{ART/}}
 \DeclareGraphicsExtensions{.pdf,.png,.jpg}

\newcommand{\nc}{\newcommand}

\newcommand{\esm}{\ensuremath}
\newcommand{\mcl}{\multicolumn}
\newcommand{\sspc}[1]{#1\ }
\nc{\etal}{\sspc{et~al.}}
\nc{\eg}{\sspc{e.g.}}
\nc{\etc}{\sspc{etc.}}
\nc{\cf}{\sspc{cf.}}
\nc{\ie}{\sspc{i.e.}}
\nc{\cfig}[1]{\centerline{#1}}
\nc{\Ncol}[1]{\mcl{#1}{c}{~}}
\nc{\mcn}[1]{\mcl{#1}{l}{~}}
\nc{\mcN}[2]{\mcl{#1}{c}{#2}}
\nc{\mcc}[1]{\mcl{1}{c}{#1}}
 \nc{\AAA}{\sspc{\esm{\lambda\lambda}}}
 \nc{\amm}{\sspc{\,\AA\,mm$^{-1}$}}
 \nc{\kms}{\sspc{\,\esm{\text{km s}^{-1}}}}
 \nc{\msun}{\sspc{\,\esm{\text{M}_{\odot}}}}
 \nc{\rsun}{\sspc{\,\esm{\text{R}_{\odot}}}}
 \nc{\lsun}{\sspc{\,\esm{\text{L}_{\odot}}}}
 \nc{\yr}{\sspc{\,\esm{\text{yr}}}}
 \nc{\kpc}{\sspc{\,\esm{\text{kpc}}}}
 \nc{\halpha}{\sspc{\,\mbox{H$\alpha$}}}
 \nc{\hii}{\rm H\,{\sc ii} }
 \nc{\ion}[2]{\sspc{#1\,{\scshape#2}}}
 \nc{\dgr}{\esm{^\circ}}
 \nc{\asec}{\sspc{\,\raisebox{0.25ex}{\scshape"}}}
 \nc{\Mag}[2]{#1\fm#2}
 \nc{\dex}[1]{\esm{10^{#1}}}
 \nc{\tdex}[1]{\esm{\times10^{#1}}}
 \nc{\todo}[1]{\sspc{(\textbf{TODO:} #1)}}
 \nc{\obj}[1]{\sspc{#1}}
 \nc{\rotsed}{\sspc{ROTSE--IIId}}
 \nc{\rotse}{\sspc{ROTSE--III}}
\begin{document}
\title{%
  Astronomical Site Selection for Turkey Using GIS Techniques%
}
\titlerunning{%
  Site Selection for Turkey%
}
\author{%
N.\      Aksaker \and
S.\,K.\  Yerli \and
M.\,A.\  Erdo\u{g}an \and
E.\      Erdi \and
K.\      Kaba \and
T.\      Ak \and
Z.\      Aslan \and
V.\      Bak{\i}\c{s} \and
O.\      Demircan \and
S.\      Evren \and
V.\      Keskin \and
\.I.\    K\"u\c{c}\"uk \and
T.\      \"Ozdemir \and
T.\      \"Oz{\i}\c{s}{\i}k \and
S.\,O.\  Selam%
}
\authorrunning{%
  N.Aksaker \etal%
}
\institute{%
  N.\      Aksaker \and M.\,A.\  Erdo\u{g}an
  \at \c{C}ukurova University, Vocational School of Technical Sciences, Adana, Turkey
  \and
  S.\,K.\  Yerli
  \at Orta Do\u{g}u Teknik \"Universitesi, Department of Physics, Ankara, Turkey
  \mailname yerli@metu.edu.tr
  \and
  E.\      Erdi
  \at Remote Sensing Division, Turkish State Meteorological Service, Ankara,Turkey
  \and
  K.\      Kaba
  \at \c{C}ukurova University, Department of Physics Adana, Turkey
  \and
  T.\      Ak
  \at \.Istanbul University, Faculty of Science, Department of Astronomy and Space Sciences, \.Istanbul, Turkey
  \and
  Z.\      Aslan
  \at 1427. Cd., No:4/24, Balgat, Ankara, Turkey
  \and
  V.\      Bak{\i}\c{s}
  \at Akdeniz University, Science Faculty, Space Sciences and Technologies Department, Antalya, Turkey
  \and
  O.\      Demircan
  \at \c{C}anakkale Onsekiz Mart University, Science and Letter Faculty, Space Sciences and Technologies Department, \c{C}anakkale, Turkey
  \and
  S.\      Evren \and V.\      Keskin
  \at Ege University, Science Faculty, Astronomy and Space Sciences Department, \.Izmir, Turkey
  \and
  \.I.\    K\"u\c{c}\"uk
  \at Erciyes University, Faculty of Science, Department of Astronomy and Space Sciences, Kayseri, Turkey
  \and
  T.\      \"Ozdemir
  \at \.Inönü Univerity, Science and Letter Faculty, Physics Department, Malatya, Turkey
  \and
  T.\      \"Oz{\i}\c{s}{\i}k
  \at T\"UB\.ITAK National Observatory, Antalya, Turkey
  \and
  S.\,O.\  Selam
  \at Ankara University, Faculty of Science, Department of Astronomy and Space Sciences, Ankara, Turkey%
}
\date{%
	Received: date / Accepted: date%
}

\maketitle
\begin{abstract}
A site selection of potential observatory locations in Turkey have been carried out by using Multi-Criteria Decision Analysis (MCDA) coupled with Geographical Information Systems (GIS) and satellite imagery which in turn reduced cost and time and increased the  accuracy of the final outcome.
The layers of cloud cover, digital elevation model, artificial lights, precipitable water vapor, aerosol optical thickness and wind speed were studied in the GIS system. In conclusion of MCDA, the most suitable regions were found to be located in a strip crossing from southwest to northeast including also a diverted region in southeast of Turkey. These regions are thus our prime candidate locations for future on-site testing. In addition to this major outcome, this study has also been applied to locations of major observatories sites. Since no goal is set for \textit{the best}, the results of this study is limited with a list of positions. Therefore, the list has to be further confirmed with on-site tests. A national funding has been awarded to produce a prototype of an on-site test unit (to measure both astronomical and meteorological parameters) which might be used in this list of locations.
\end{abstract}

\PACS{%
(PACS codes)%
}
\section{Introduction}
\label{s:intro}

Selecting an astronomical site is always an important part of progress of astronomy in which one seeks the most suitable place for projects and experiments where ideas and dreams will be fulfilled. Therefore, it will involve a large amount of funding to procure the job. Achievements in both mechanical and optical engineering, and material sciences help institutions to go for relatively cost-efficient projects of building dedicated telescopes and observatories. Even though the total cost decreases, they have to be still operated cost-efficiently from astronomical point of view \ie locate them on \textit{sites} where annual observing rate is maximized. Therefore, finding a good site for an astronomical observatory becomes as important as funding the observatory itself.

\begin{figure*}
\cfig{\includegraphics[width=\textwidth]{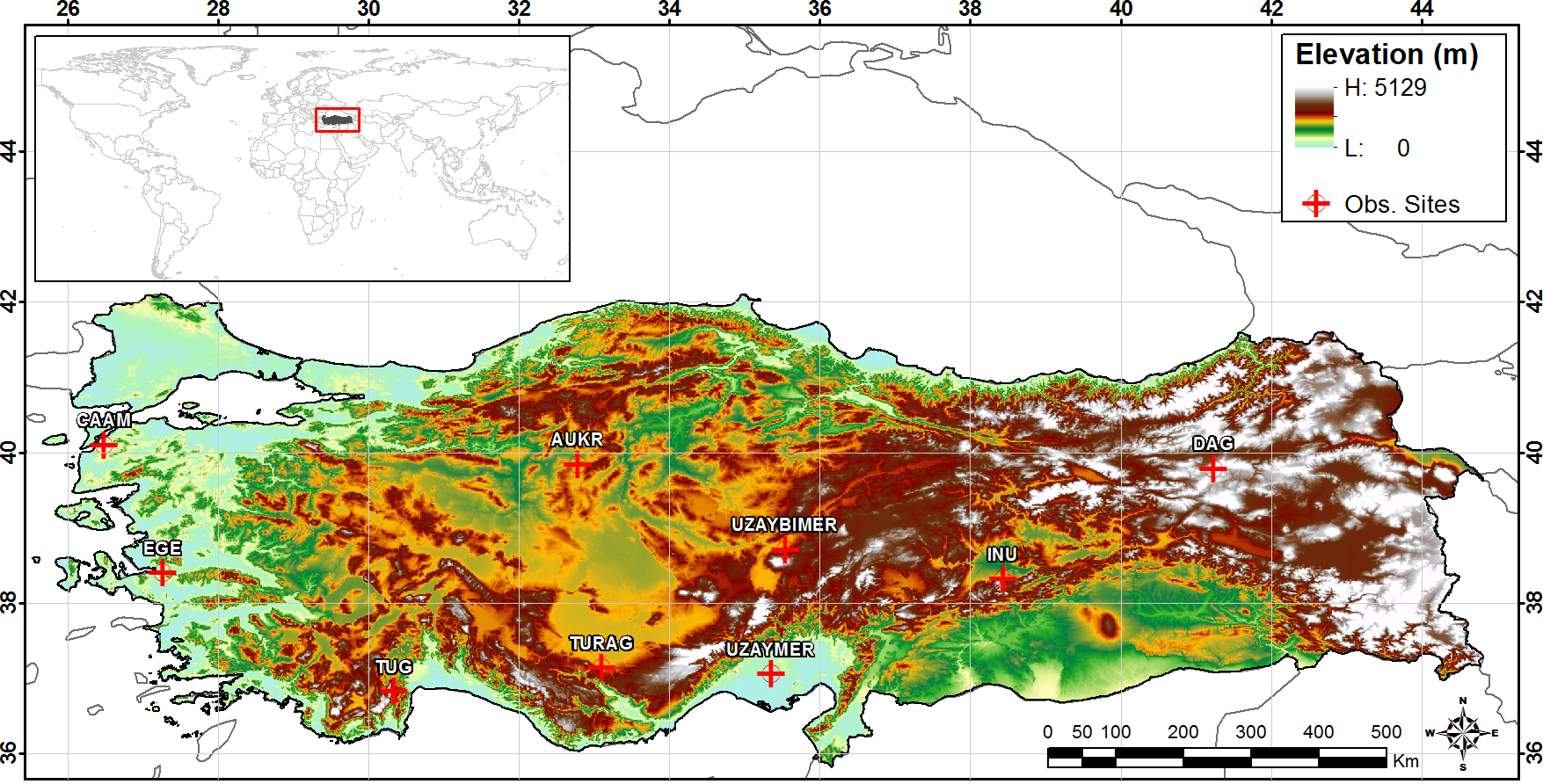}}
\caption{General view of Turkey derived from Digital Elevation Model.
Institutional observatory sites are marked with a cross hair and their abbreviations are given in Table \ref{t:siteinf}.}
\label{f:world}
\end{figure*}

To place expensive and delicate astronomical instruments, one usually require a land surface which is well above the sea level (to collect more and more undisturbed light from above the atmosphere) and well away from human activities (to increase the quality of the collected light). Surface area of geographic locations above \eg 2000 m is just 32\% of Earth's terrestrial surface. The same area is 12\% for Turkey; pushing the country to a favorable place. Moreover, distribution of geological structures will limit our search to southern, south-eastern, eastern part of Turkey (see Figure \ref{f:world}). However, when the list of largest telescopes%
\footnote{for 2014: \url{http://en.wikipedia.org/wiki/List_of_largest_optical_reflecting_telescopes}}
are considered there are two obvious gaps in \textit{telescope coverage} around the world: (1) longitude wise (continuity of an observing run starting from East is broken at around 30-45 degrees East);
(2) latitude wise (dedicated surveys when coupled with longitudinal coverage might have difficulties in covering both hemispheres at the same time with large telescopes; see \eg the vision of collaboration between hemispheres in ``ESO in 2020s'' workshop).
Therefore with this study we also attempt to fill these gap by providing potential sites in Turkey for large telescopes.

Furthermore, our aim is not limited to just finding the best observing site but also to construct a list of potential observatory sites with different goals and dedications. However, this list will be effected mostly with demographic changes of the human activity. As the development of countries and spread of large cities reach to an enormous level, finding good sites for astronomical observatories becomes more and more difficult, as well as essential. Thus, a good astronomical observatory site must be protected from the disruptive factors (\eg city lights, industrial pollution, near settlement of people, mining activities), making the site selection a very complicated and time consuming job with many parameters and/or criteria.

``Searching'' for astronomical observatory sites throughout the whole Earth surface has already been done for ELT (Extremely Large Telescope) and TMT (Thirty Meter Telescope) using some automated software with data sets ranging from satellites to meteorological databases (\citealp{2006Msngr.125...44S,2008SPIE.7012E..1YG}). The idea behind these particular works will not fit to \textit{other} search attempts because they were looking for \textbf{the best} location on ``Earth'' so that they could get above 96\% annual observing rate (corresponding to losing at most two weeks)
with the best weather and atmospheric conditions --including  high angular resolution parameters-- for astronomy.
Thus, they have reconfirmed basics of site selection: atmospheric activity at the location and within its surroundings should be both minimum and stable (high-dry-calm). Rest of the criteria will naturally rely on the cost-efficiency of the project (maintenance, infra-structure, logistic including geophysical properties and seismicity, access, altitude, \etc  and cultural, social and policies issues).

These site selection parameters and criteria have to be applied for a large surface area to come up with a list of potential locations. In itself, the whole process has to be automated, however, requiring accurate and adequate data sets. An example of selection of an ideal site using GIS (Geographical Information Systems) and astroclimatology layers (cloud cover, atmospheric humidity, aerosol content, air temperature, airflow direction, strength and turbulence) was successfully made by \cite{2004SPIE.5489..102G}. They worked on a case study with 11 layers including topography, total cloud cover and precipitable water vapor for ALMA (The Atacama Large Millimeter/submillimeter Array) project. As a result they found a site in Chajnantor, Chile as the best place for radio astronomy (\citealp{2006Msngr.125...44S}).

Due to its complexity, site selection involves evaluation of multiple criteria from different data sets (for 11 factors see \eg \citealp{2010NST}) and can only be resolved by dedicated analysis methods (\eg Multi-Criteria Decision Analysis - MCDA; \citealp{2005JW, 2005S}) using structured data of GIS. GIS offers to the user an efficient, cost effective and viable data manipulations, results and solutions in a considerably fast time spans where they can be successfully applied many times (\citealp{2013PASA...30....2H, 2005MeApp..12...77G}).
\begin{table*}
\centering
\caption{Major institutional observatory sites in Turkey. The list is ordered from West to East with respect to longitude of the site location. Locations are directly copied from published values and elevation of the site is taken from the DEM model explained in \S\ref{subsec:ele}.}
\label{t:siteinf}
\begin{tabular}{@{}l@{~}c@{~}@{~}c@{~}c@{~}c@{~}r@{}}
\hline
Observatory Site
& City
& Longitude
& Latitude
& Elevation
\\

&
& (\dgr~ East)
& (\dgr~ North)
& (m)
\\
\hline
\c{C}AAM -- \c{C}anakkale Astrophysics Research Center
	& \c{C}anakkale
	& 26.48
	& 40.01
	& 373\\
EGE -- Ege University Observatory
	& \.Izmir
	& 27.27
	& 38.40
	& 622\\
TUG -- T\"UB\.ITAK National Observatory
	& Antalya
	& 30.34
	& 36.82
	& 2436\\
A\"UKR -- Ankara University Kreiken Observatory
	& Ankara
	& 32.78
	& 39.84
	& 1254\\
TURAG -- Turkish National Radio Astronomy Observatory Site
	& Karaman
	& 33.09
	& 37.14
	& 1062\\
UZAYMER -- Uzay Bil. ve G\"une\c{s} En. Ar\c{s}. Uyg. Mrk.
	& Adana
	& 35.35
	& 37.06
	& 112\\
UZAYB\.IMER -- Astr. ve Uzay Bil. G\"oz. Uyg. ve Ar\c{s}. Mrk.
	& Kayseri
	& 35.55
	& 38.71
	& 1094\\
\.IN\"U - \.In\"on\"u University Observatory
	& Malatya
	& 38.44
	& 38.32
	& 1021\\
DAG -- Do\u{g}u Anadolu G\"ozlemevi
	& Erzurum
	& 41.23
	& 39.78
	& 3102\\
\hline
\end{tabular}
\end{table*}

Historically, within demographic boundaries of Turkey, \cite{1989A&A...208..385A} had showed using long-term meteorological records, on-site observations and visual inspection on METEOSAT and NOAA-7 satellite images that southwest and southeast of Turkey contain good potential observatory sites. They found that \textit{Bak{\i}rl{\i}tepe, Antalya, Turkey} compares very favorably at the time of survey duration with the Roque de los Muchachos Observatory on La Palma, which is among the world's best sites.

Recently a regional site selection was made by \cite{2013AdSpR..52...39K} using Multi-Criteria Decision Analysis coupled with Geographical Information Systems (GIS-MCDA) and remote sensing technologies for Antalya province of Turkey. Their results revealed that regionally, suitable areas are located extensively in western and eastern part of Antalya.

The Table \ref{t:siteinf} gives details of all major institutional observatory sites in Turkey including the one that will accommodate a 4 m telescope in around 2019, the largest in Turkey \ie DAG (Do\u{g}u Anadolu G\"ozlemevi - East Anatolian Observatory). It has to be noted that even though site location of TURAG has been included \citealp{2012ExA....33....1K} there are no institutional facilities established yet. However, in the scope of this work, it will be counted as an institutional site.

The primary goal of this study is to make a \textit{list} of potential \textit{new} observing sites in Turkey using GIS-MCDA so that in future, both extended GIS and further on-site studies could easily be carried out without requiring manipulation of large GIS data sets.

\subsection{A brief introduction to Turkey's geography and climatology}

Turkey is located in Eurasia continent between southeastern Europe and
southwestern Asia at which these lands are connected via Bosporus strait.
It borders to the Black Sea, between Bulgaria and Georgia; the Aegean Sea and
the Mediterranean Sea, between Greece and Syria.
The country has land borders to Armenia, Azerbaijan, Bulgaria, Georgia, Greece,
Iran, Iraq, Syria amounting to 2648 kilometers whereas the coastline is around
7200 kilometers.
The main terrain structure contains a high central plateau (Anatolia) having
narrow coastal plain with several mountain ranges.
Geographic coordinates ranges in between around 26--25 East and 36-42 North
degrees.

Turkey's coastal climate has two variations: bordering to Black Sea the climate
is temperate Oceanic with warm, wet summers and cool to cold wet winters;
bordering to Aegean Sea and the Mediterranean Sea, a typical Mediterranean
climate is observed with hot, dry summers and mild to cool, wet winters.
Due to a high mountain ridge ranging along with the Black Sea coastline, the
area receives the greatest amount of precipitation.
Snow coverage is rare in most of the coastal areas of the Aegean Sea and the
Mediterranean Sea.
Climatic conditions is much harsher in interior land areas.
Mountain ridges along with both coastlines of the Black Sea and the
Mediterranean Sea prevent maritime influences from extending inland which gives
the Anatolian plateau a climate with contrasting seasonal changes.
Winters on the plateau, however, are especially severe with temperatures
reaching to $-30$ to $-40$ \dgr C.
Similarly, summers are hot and dry with temperatures generally above $30$
\dgr C during the day.
A more detailed climatic analysis of Turkey is given in \citealp{mgm2008}.

\subsection{Summary of Sections}

Datasets and layer definitions are given in \S\ref{sec:obsdat}.
Both data and MCDA analysis are given in \S\ref{subsec:ana}.
Results and discussion of the analysis is given in \S\ref{sec:results} and
conclusion is given in \S\ref{sec:conc}.
\section{Data Sets of Layers}
\label{sec:obsdat}

We have started with 11 layers (CC: Cloud Coverage, DEM: Digital Elevation Model, AL: Artificial Light, PWV: Precipitable Water Vapor, AOD: Aerosol Optical Depth, WS: Wind Speed, geologic structure, seismicity, distance to fault lines, distance to settlements and mining activity). Since most of these layers were publicly available they were acquired and studied with standard procedures. These layers constitute the essential selection criteria to find the best possible observatory sites before in-situ detailed selection studies (\citealp{2013AdSpR..52...39K}). Therefore, since the ultimate aim of using above layers is to find a location where Z-values (\ie the value of the pixel at a geographic location) are minimum/maximum at these locations, one could easily reduce associated layers with the following reasons:
For the layers of geologic structure, seismicity, distance to fault lines, distance to settlements and mining activity; the layer's effect can only be achieved with on-site observations unless very expensive data sets are used. Therefore, adding the effect of the layer at the GIS level will have little effect on the major \textit{astronomical} criteria.

Therefore, the remaining six layers (CC, DEM, AL, PWV, AOD and WS) are the most important and adequate layers to find the best astronomical observatory sites when their meteorological, geographical and anthropogenic properties are considered.

\subsection{Cloud Coverage - CC}
\label{subsec:clo}
CC (\ie Cloudiness) is the first and the most important factor in astronomical site selection as discussed by \cite{1983ESOC...17..217A}. CC data set which were obtained from MODIS (Moderate Resolution Imaging Spectroradiometer) instruments on board the TERRA and AQUA polar orbiting satellites and from geo-stationary METEOSAT satellites. The MODIS instruments are operational for global remote sensing of the land, ocean and atmosphere (\citealp{2004AdSpR..34..710S}).
It acquires data in 36 high spectral resolution bands between 0.415 and 14.235 $\mu$ with a spatial resolutions of 250 m (2 bands), 500 m (5 bands), and 1000 m (29 bands). Frequency of satellite (AQUA and TERRA) passages over Turkey is four times in a day (2 for night time and 2 for day time). CC data sets that we have used were taken from Cloud Mask product, namely MYD35/MOD35 (\citealp{2007HydPr..21.1534H, 2004AdSpR..34..710S}).
The MODIS Level 2 data set for 2003-2012 were downloaded from the MODIS web page\footnote{\url{http://modis-atmos.gsfc.nasa.gov/}}.

On the other hand METEOSAT satellites provide main full disk imagery service over Europe and Africa (with image acquisition frequency of 15 minutes). SEVIRI sensor on board the satellite has 12 bands covering spectral range from 0.6 $\mu$m to 13.4 $\mu$m and with spatial resolutions of 3 km and 1 km (for only channel 12) at Nadir (\citealp{2002BAMS...83..977S}). CC product generated in \textit{Meteorological Products Extraction Facility} of EUMETSAT has a complex algorithm using visible and infrared bands in addition to the atmospheric profiles of \textit{Numerical Weather Prediction} models. Note that, since cloud product of both satellites is designed to be used only for land surfaces, pixels close to boundaries between sea and land surface are not calculated properly due to inefficiency of cloud product algorithm where a narrow strip might be introduced close to the sea boundaries(see top-left panel of Figure \ref{f:layers}). These data sets have been obtained through \textit{Turkish State Meteorological Service} where the cloud mask products are retrieved regularly from EUMETSAT Archive.

Both cloud mask products have their own advantages and disadvantages. MODIS has high spatial resolution (1 km) but very low temporal resolution (2 passes per night), whereas METEOSAT has low spatial resolution (around 5 km for Turkey) but high temporal resolution of 15 minutes. Therefore, METEOSAT Cloud Mask product has been chosen due to its time resolution so that night time coverage can be achieved much better.
Note that this choice is purely from the astronomical point of view. It tries to eliminate locations having large night-time cloud coverage by leaving determination of correct location to on-site tests if the location has already been chosen as favorable for astronomy.
The time span is restricted to last 5 years to achieve fast computing time; as a future work the full coverage span will be studied elsewhere.

\subsection{Digital Elevation Model - DEM}
\label{subsec:ele}
DEM is a scale of elevation extracted from remotely sensed data.
Elevation (and therefore DEM layer) is another important factor for site selection because atmospheric thickness decreases as elevation increases and therefore the sites are less effected from tropospheric activities. Thus, ideally, observatories are constructed above the atmospheric inversion layers%
\footnote{From the astronomical point of view, inversion layer is defined as the elevation where convective motion of warm air is trapped below where the temperature gradient with respect to elevation above the layer remains negative however below the layer it becomes positive.}.
 A recent DEM data set with a spatial resolution of 30 m is used for Turkey.

The data for this layer were obtained from ASTER (\textit{Advanced Spaceborne Thermal Emission and Reflection Radiometer}) GDEM (\textit{Global DEM}). The Ministry of Economy, Trade and Industry of Japan (METI) and the National Aeronautics and Space Administration (NASA) were collaborated to develop an elevation model for ASTER GDEM where the coverage was the whole Earth surface. ASTER GDEM data for 2012 can be downloaded free of charge from their web site\footnote{\url{http://gdem.ersdac.jspacesystems.or.jp/}}.

\subsection{Artificial Light - AL}
\label{subsec:lig}
AL is due to human activity seen from satellites.
Quality of astronomical observations are directly affected by the light pollution around the site due to decrease in sky's natural magnitude limit by introducing artificial luminosity in various part of wavelength coverage. If the intensity of sky background increases then number of observable sources dramatically decreases. So, astronomical sites should be selected among locations having lower sky brightness values. Therefore, we have used nighttime imagery of stable anthropogenic lights (\ie cleaned from \eg forest fires and lightings) obtained from the DMSP/OLS (\textit{Defense Meteorological Satellite Program's Operational Linescan System}) which offers detailed and up-to-date (2012) settlement light maps. These maps can be used as a source of light pollution.

\subsection{Precipitable Water Vapor - PWV}
\label{subsec:pwv}
PWV is the total column of water vapor in Zenith direction (\citealp{Ferrare2002}). It is the main contributor to the opacity in infrared wavelengths of atmospheric window. In the context of astronomical observations PWV also means an infrared activity above an observatory site (\citealp{2010MNRAS.405.2683G}). Therefore it has to stay at minimum levels to achieve successful scientific goals, especially in infrared astronomy. For this reason telescopes with infrared capabilities are usually placed on dry, stable and high altitude plateaus. The PWV content in the atmosphere can be measured from satellite imagery. We have used MODIS AQUA satellite's Water Vapor Infrared (MYD05) data.

\subsection{Aerosol - AOD}
\label{subsec:ars}
Aerosol Optical Depth (AOD) is the level of change of atmospheric transparency in daylight due to aerosols.
The AOD available in this project was also believed to be related to atmospheric extinction however it is still debated (\citealp{2004SPIE.5489..138S, 2004SPIE.5571..105V}). To obtain this layer we have used MODIS's Corrected Optical Depth Land (MYD04) product. Optical Depth layer was generally used in studies of transparency of the atmosphere (\citealp{2004SPIE.5489..102G}). As is very well known, measurements of AOD using satellite imagery in the atmosphere could only be sensed from the daylight frames.

\subsection{Wind Speed - WS}
\label{subsec:wnd}
Astronomical seeing is related with the WS distribution right above the site locations (\citealp{2010RAA....10.1061L}) and with the velocity of the turbulence in the atmosphere (\citealp{2002ESOC...58..321S}, \citealp{2005MNRAS.356..849G}, \citealp{2006SPIE.6267E..1XV}). Since the behavior of the wind varies with both location and altitude, constructing the layer was not trivial. To accommodate a practical work around, daily means of the WS measurements taken from Turkish State Meteorological Service\footnote{\url{http://www.mgm.gov.tr/}} of 362 stations spread around the boundaries of Turkey for over 40 years (upto 2012) were taken and converted to GIS compatible format so that the study area can be studied as a single layer.

It has to be noted also that there were two main issues for the integrity of the layer: (1) weather stations were located mainly in or around city locations having DEM values well below 2000 m; (2) since wind speed usually increases as elevation increases, collected station data might not be representing the sites that we are looking for.

\subsection{Layer Descriptions}

\begin{figure*}
\cfig{%
\includegraphics[width=0.45\textwidth]{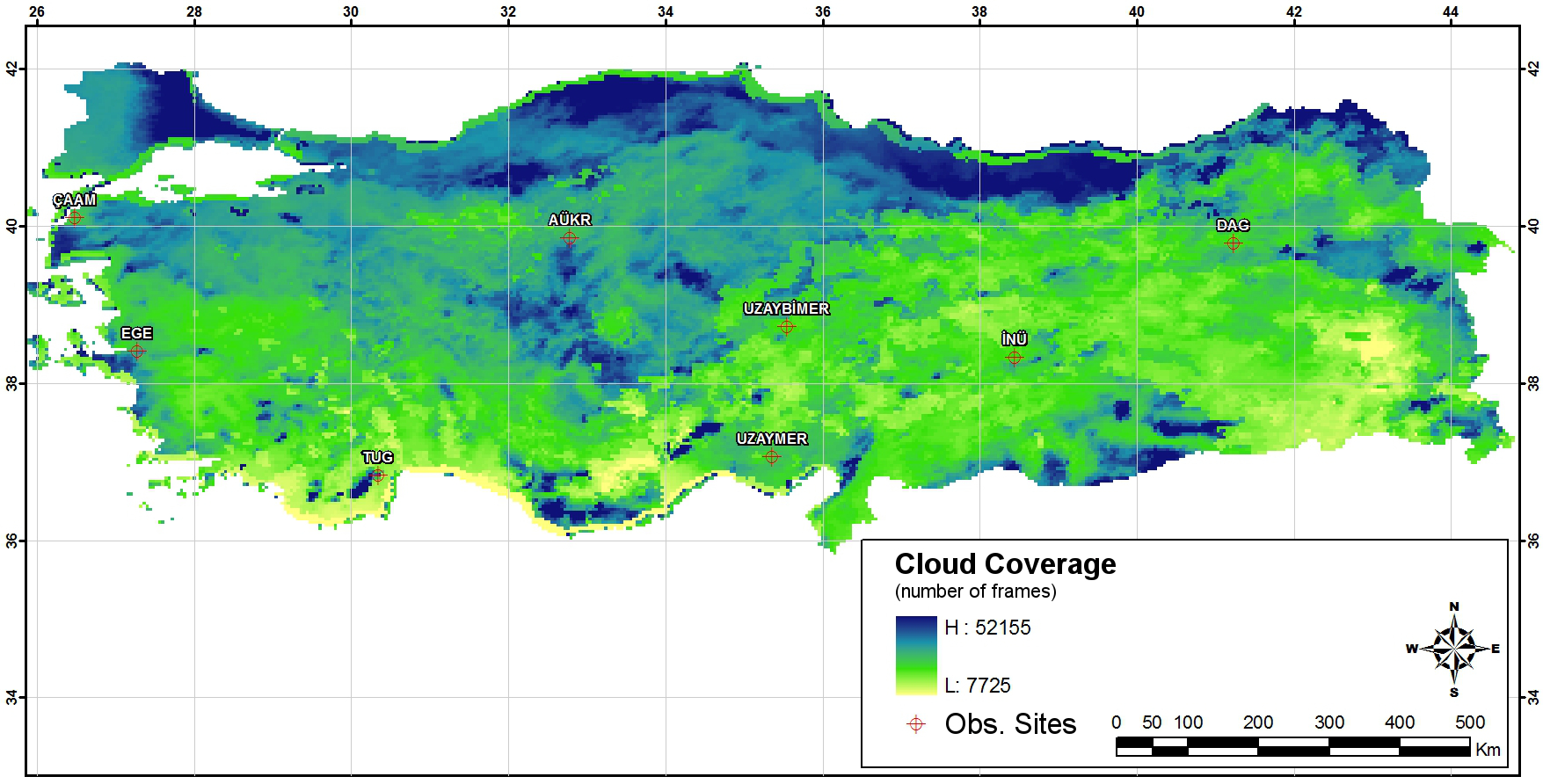}
\includegraphics[width=0.45\textwidth]{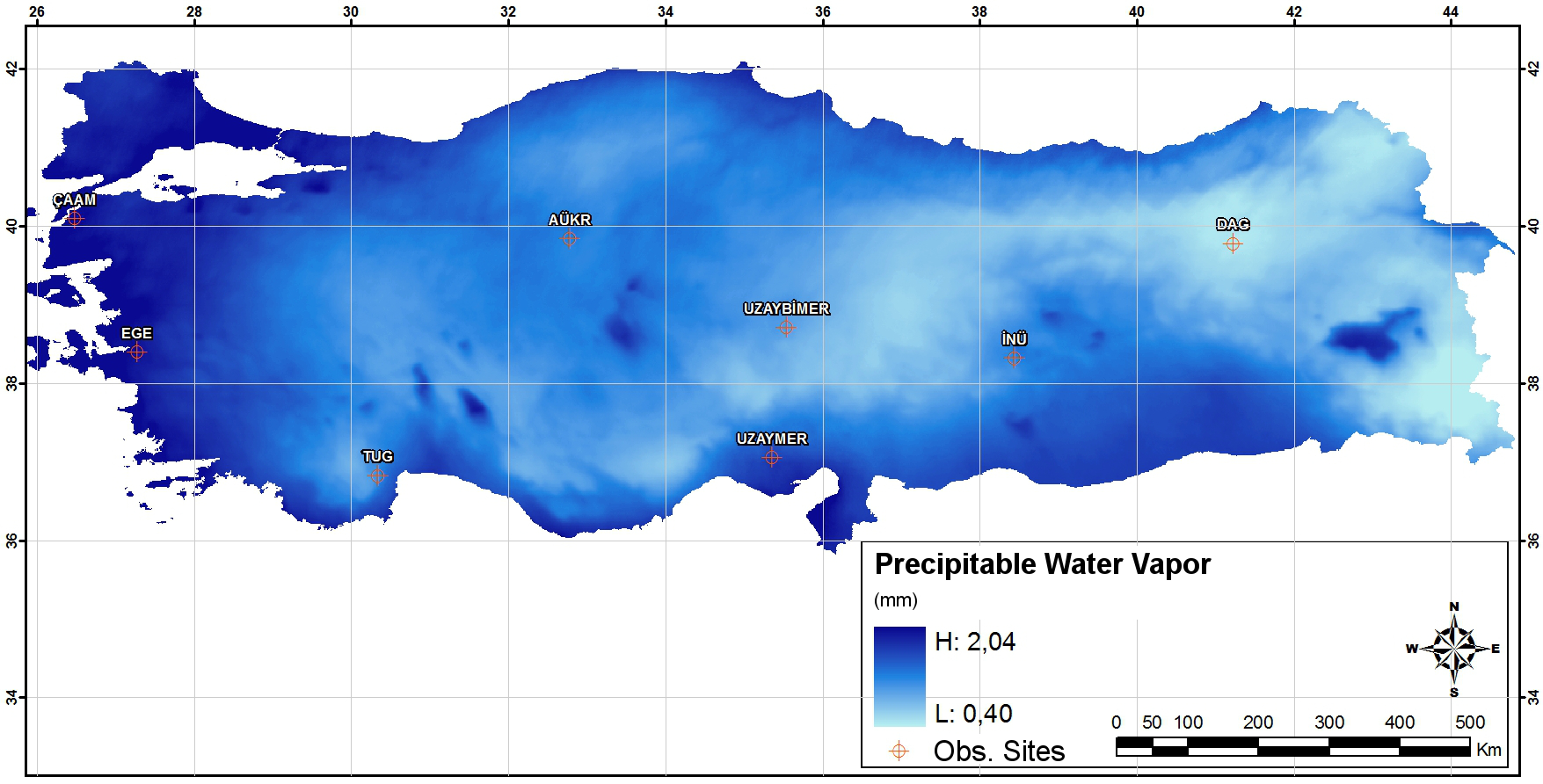}}
\cfig{%
\includegraphics[width=0.45\textwidth]{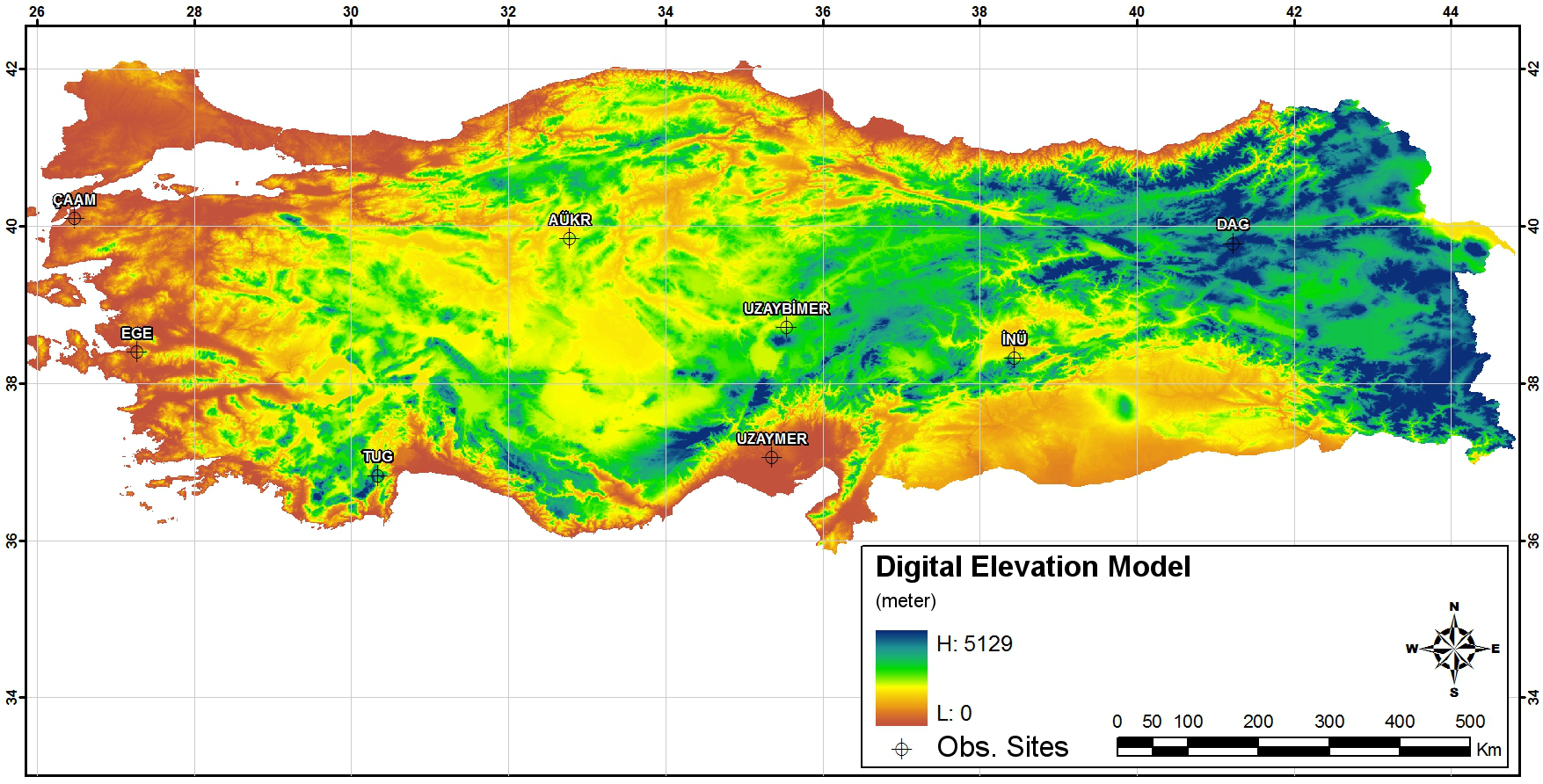}
\includegraphics[width=0.45\textwidth]{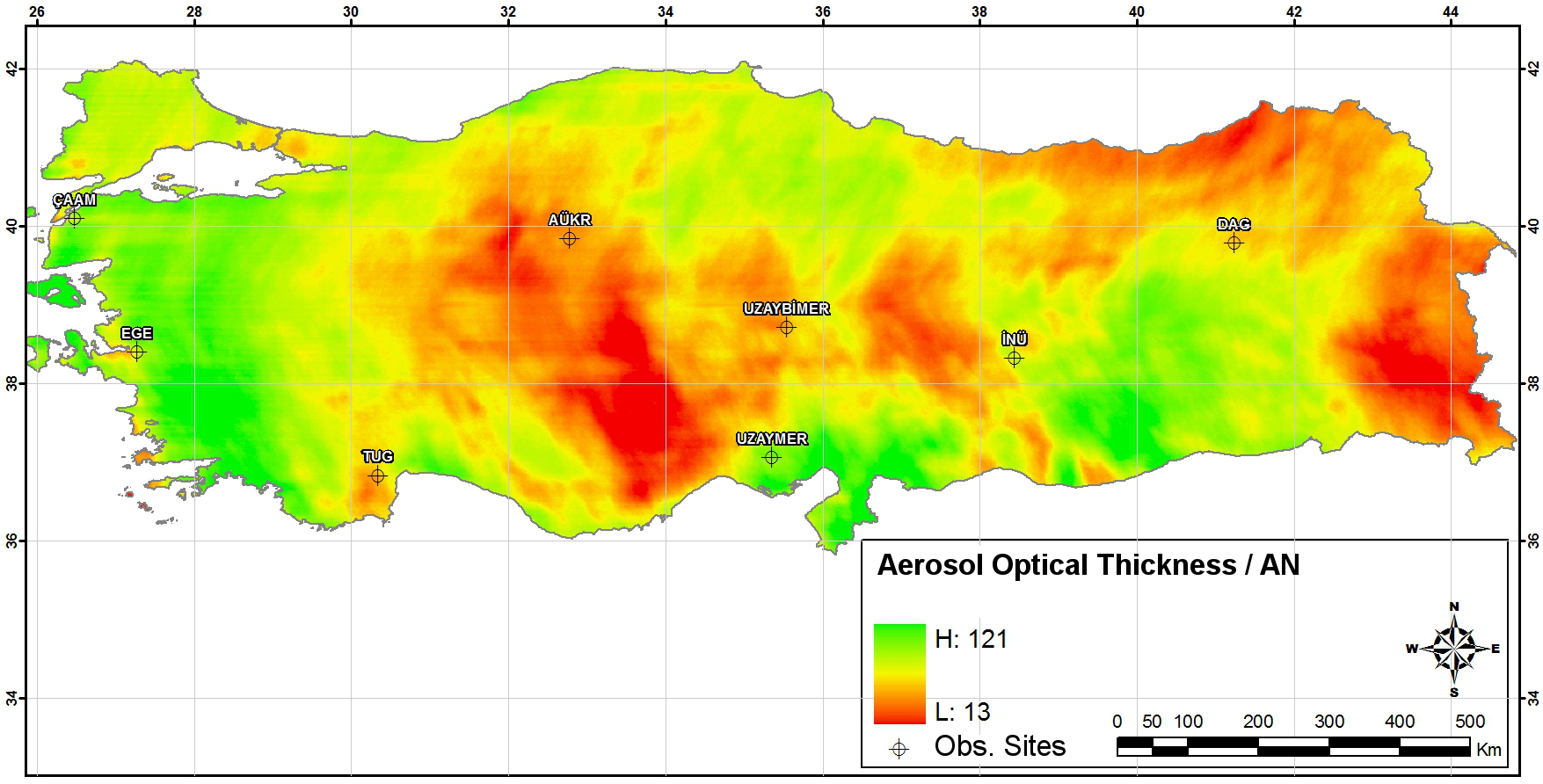}}
\cfig{%
\includegraphics[width=0.45\textwidth]{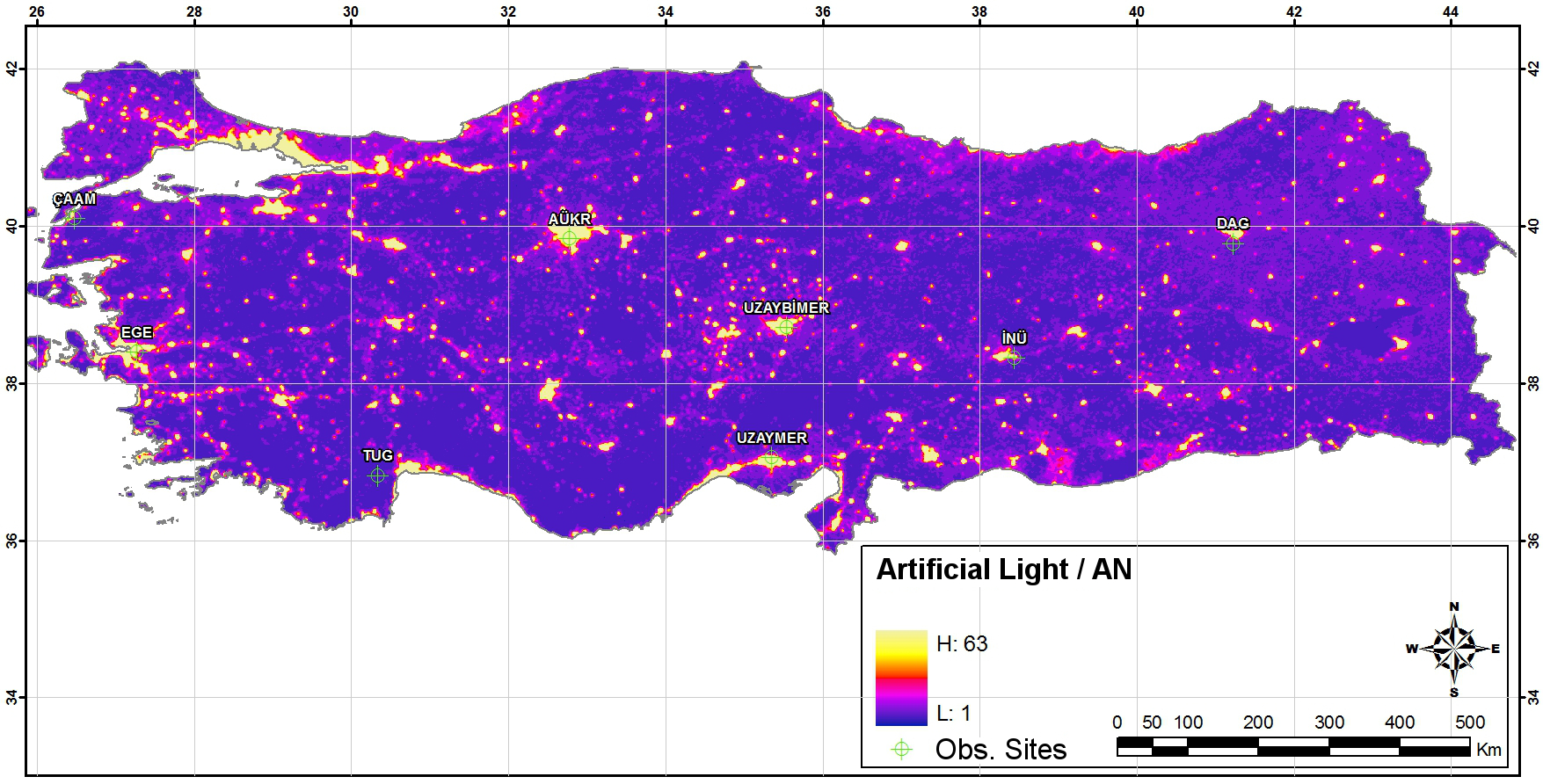}
\includegraphics[width=0.45\textwidth]{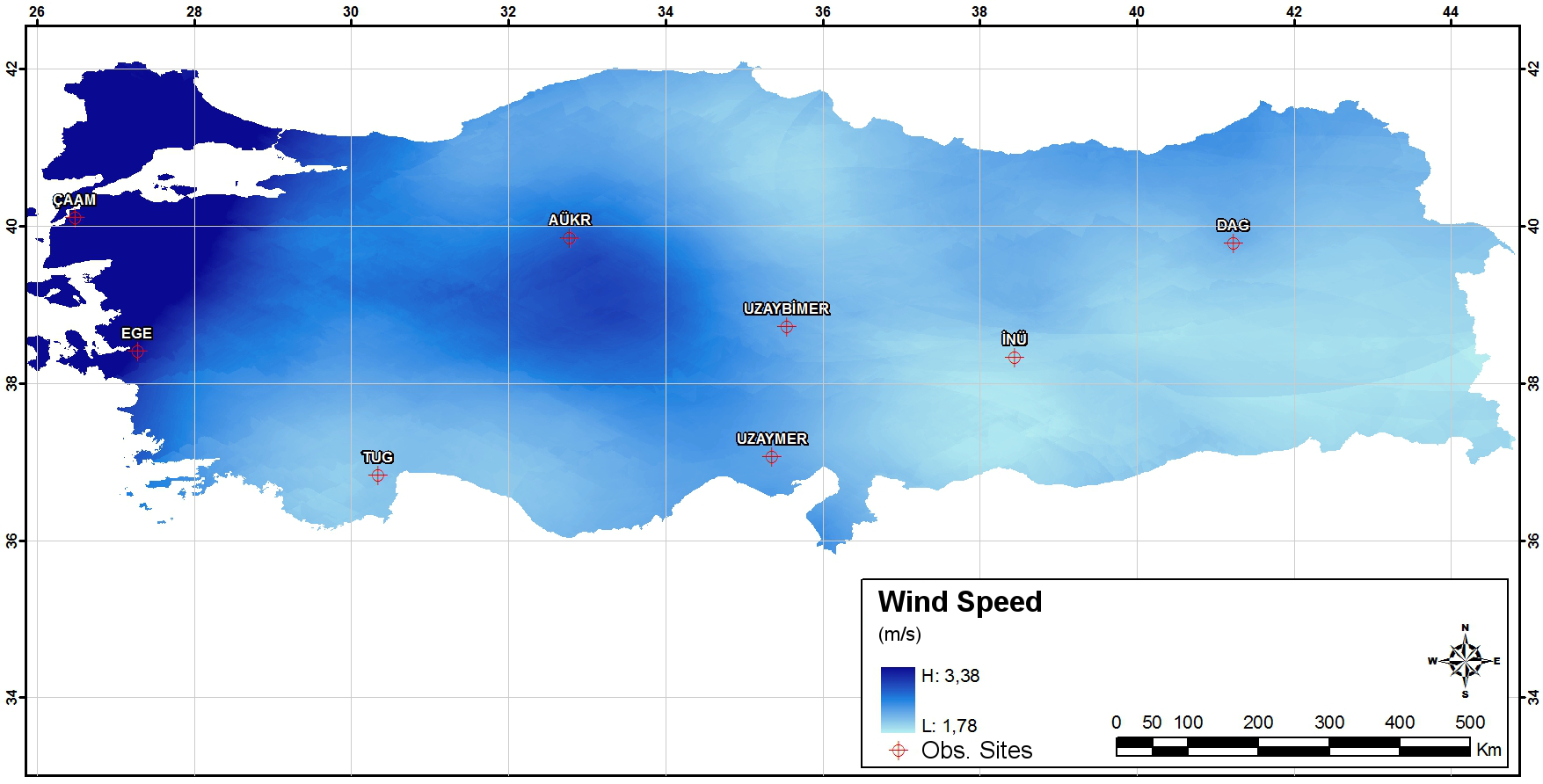}}
\caption{Layers used in MCDA.
Units of the layers are as follows. CC: \textit{number of cloudy frames};
DEM: \textit{meters};
PWV: \textit{mm};
WS: \textit{m/s};
The rest: \textit{arbitrary numbers} (\ie AN).
Other layer descriptions are given in \S\ref{sec:obsdat}.}
\label{f:layers}
\end{figure*}

The abovementioned layers are shown in Figure \ref{f:layers}. The coloring of the layer was done according to each layer's own data limits \ie data between minimum and maximum values (in arbitrary units) are gray scaled between two chosen colors. On top of each layer we have also located the positions of the major institutional observatory sites (both current and future ones; see Table \ref{t:siteinf}). The coloring was trimmed at the Turkish demographic borders.

\section{Analysis}
\label{subsec:ana}

The analysis methods of layers from different sources having different data types are explained one by one.

We firstly started with the \textit{METEOSAT Cloud Mask}. This product is ordered from EUMETSAT archive for the years between 2008 and 2012; and then filtered for the times between 20:00 to 03:00 UTC for every 15 minutes in GRIB2 file format; approximately 52000 GRIB2 files. They were decoded, re-projected and subsetted for Turkish domain and converted to PNG image format with an in house developed C code. Then, CC for every pixel was calculated from those processed files to form \textit{the CC raster layer} for GIS environment.

Elevation data are derived from 121 frames of ASTER GDEM. These frames are combined to a single raster layer by mosaicking. Thirty meter spatial resolution of DEM is resampled to 100 m to reduce computing time.

DMSP/OLS up-to-date light map for 2012 is downloaded from their web page\footnote{\url{http://ngdc.noaa.gov/eog/}}. The night lights raster data are produced from cloud-free composites made using all available archived DMSP-OLS smooth resolution data. The product is composed of 30 arcsecond grids, spanning -180 to 180 degrees longitude and -65 to 75 degrees latitude. This raster is extracted according to the extends of the study area.

There were about ~5000 PWV images obtained from MODIS archive near the midnight passage of the satellite. This product is downloaded from MODIS archive for the years between 2003 and 2012 in HDF format covering only the study area boundaries. There exist one or two images for each night. When a particular image fails to cover whole study area, several passages had to be combined and converted to the GeoTIFF format using a software called \textit{HDF-EOS To GeoTIFF Conversion Tool (HEG)}. The real PWV values were calculated using in house developed IDL routines with a known conversion factor which were in the header of each image.

MODIS MYD04 data (\ie optical depth) were used to create the AOD layer from 2682 images. The analysis used in PWV was also used for the \textit{optical depth} layer.

As is explained in the previous sections, the WS layer data are constructed from Turkish State Meteorological Service archives. These huge data sets were collected from almost randomly distributed stations and therefore they firstly had to be interpolated using the Krigging model (see \citealp{2008IJCli..28..947L} for a similar study) and then they were averaged over its time span of 40 years to produce a single value for each position in the study area. Finally, it was trimmed from Turkish border.

All layer data were then georeferenced to a common coordinate system called \textit{Geographic Coordinate System} (GCS) with a datum of WGS84 and the base resolution is downgraded to 4-5 km for each layer.

The final criterion in constructing a combined map of layers are given below. Depending on each layer's astronomical suitability, layer values are rescaled between 0 (minimum - worse site location) and 1 (maximum - best site location) to prepare the layer for the main analysis method (see the next section for the details):
\begin{itemize}

\item \textbf{Quality of the site increases as the CC decreases.} Therefore the layer is scaled inversely: the highest number of CC was taken as minimum.

However, CC layer had to be worked out according to the nature of Anatolian atmospheric conditions. The CC is usually increased during the seasonal changes where they occur twice a year (\ie spring and autumn).
The stability and the elevation of the inversion layer of the location during these periods are disturbed and elevation of the inversion layer tends to increase which also increases the CC of the location and leading to a biased data set \ie CC will never have a zero value.
Therefore, a simple 10\% of the whole year (\ie 52 weeks) can be taken as a single seasonal change period which corresponds to around 5 weeks. However, since potential astronomical sites are expected to be located well above average inversion layer altitude, this duration can be further decreased to 4 weeks per change.

Therefore, the data for this 8 weeks period have to be taken as the lower limit (\ie the duration is scaled to whole data set span of between 2008 and 2012 which gives 561/1801 days $\simeq$ 0.31). This effect can easily be noticed on the results of the analysis (see Figure \ref{f:casefigtr}).

\item \textbf{Higher the altitude, better the site location.} The site's altitude is expected to be higher than the inversion layer of that particular location. Certain type of vegetation stops growing at around 2000 m which is usually taken as an average value of the inversion layer around the world. Therefore a simple mask that filters out elevations below 2000 m will be given as an representative layout of potential regions for observatory sites.

\item  \textbf{Quality of the site increases as the PWV decreases.} Therefore  the layer is scaled inversely: the highest PWV value was taken as the  minimum.

\item \textbf{Quality of the site increases as the location's AL value decreases.} Therefore the layer is scaled inversely: the most lighten up location was taken as the minimum.

\item \textbf{Quality of the site increases as the WS decreases.} Therefore the layer is scaled inversely: the highest WS was taken as the minimum.

\item \textbf{Quality of the site increases as the AOD levels decreases.} Therefore the layer is scaled inversely: the highest AOD was taken as the minimum.
\end{itemize}

\subsection{MCDA Analysis}
\label{subsec:mcd}

\textit{Multi-criteria Decision Analysis} (MCDA) gives techniques and procedures where decision mechanisms could easily be achieved among many alternative choices and/or criteria (\citealp{2006InGIS..20..703M}).
In decision making processes, one could easily expect multiple criteria with various criterion precedence where MCDA helps to find the best viable choice-alternatives.
Since GIS uses geo-spatial data sets, MCDA could easily be used in our specific aim \ie the site selection.
This new type of analysis has been defined as \textit{GIS-based multi-criteria decision analysis} (GIS-MCDA) (\citealp{1995InGIS...9..251J}).
It is processed with three main steps:
\begin{enumerate}
\item In evaluating each layer (\ie factor), it is represented as a spatial distribution which shows the suitability of the criteria.
\item The layers are then standardized.
\item Weighted sum of these layers are calculated to reach the suitability of site selection.
\end{enumerate}
As for the weighted summation procedures, the weighted linear combination of layer criterion is simply $S = \sum W_i f_i$, where $W$ is the weight, $f$ is the factor of the score in that layer (\citealp{2011RenEn..36.2554Y}).
\begin{table}
\centering
\caption{Case studies of MCDA for the chosen layers. Details of the layers are given in \S\ref{subsec:mcd}.}
\label{t:case}
\begin{tabular}{@{}cccccccc@{}}
\hline\noalign{\smallskip}
Cases & CC  & DEM & AL  & PWV & AOT & WS  & Condition\\
\hline\noalign{\smallskip}
1     & 1.0 & 1.0 & 1.0 & 1.0 & 1.0 & 1.0 & - \\
2     & 1.0 & 1.0 & 1.0 &  -  & 1.0 & 1.0 & - \\
3a    & 1.0 & 1.0 & 1.0 &  -  &  -  &  -  & - \\
3b    & 1.0 & 1.0 & 1.0 &  -  &  -  &  -  & DEM $>2000$ \\
3c    & 1.0 &  -  &  -  &  -  &  -  &  -  & DEM $<2000$ \\
      &     &     &     &     &     &     & CC  $>0.16$ \\
\hline
\end{tabular}
\end{table}

A simple methodology is aimed in applying the MCDA analysis in which three cases were introduced with weights of layers (Table \ref{t:case}):
\begin{description}
\item[\textbf{Case-1}] The test case where weights of all layers are equal.

\item[\textbf{Case-2}] Due to nature of \underline{PWV}, values of the layer is strongly correlated with the DEM (more than 0.82; please see related panels in Figure \ref{f:layers} and discussion in \cite{1990JApMe..29..665P}) even though resolution of the layer was low.
Note also that the temperature and the height of the troposphere layer can also play an important role on the PWV content emphasizing this correlation.
Therefore the weight of the layer is taken as zero. The rest of the weights remain the same.

\item[\textbf{Case-3a}] The Case-2 is further improved by eliminating some other layers. Since \underline{AOD} is daylight-related, it will obviously be un-correlated with the astronomical site selection (\eg degree of correlation between AOD and CC is around 0.05) where the choice of activity happens during the night
(see \eg \citealp{2008MNRAS.391..507V} for a similar study between AOD and atmospheric extinction).
Therefore weight of this layer was also taken as zero however the layer itself \textit{have to} be studied after the MCDA-GIS site selection.

Similar un-correlation is found in the \underline{WS} layer (\eg degree of correlation between WS and CC is around 0.25): since the values change faster than other layers and the resolution of data set wasn't enough to overcome these changes, the layer only introduced a background noise structure in the weighted sums. Therefore its weight is again taken as zero (see the discussion in \S\ref{subsec:wnd}). However, as with AOD, it has to be studied locally with on-site tests.

Weights for the rest of the layers (DEM, CC and AL) remain equally the same as the previous case.

\item[\textbf{Case-3b}]
A quick-look outcome of Case-3a.
DEM values below 2000 m are filtered out from Case-3a.

\item[\textbf{Case-3c}]
Another test-base of Case-3b.
Only CC layer is taken and DEM values above 2000 m and CC values better than typical 8 weeks are filtered out from Case-3a.
\end{description}

As a result, Case-3a becomes the most appropriate one for an observing site accommodating a good seeing, longer observing time with dark sky.

\begin{figure*}
\cfig{%
\includegraphics[width=0.45\textwidth]{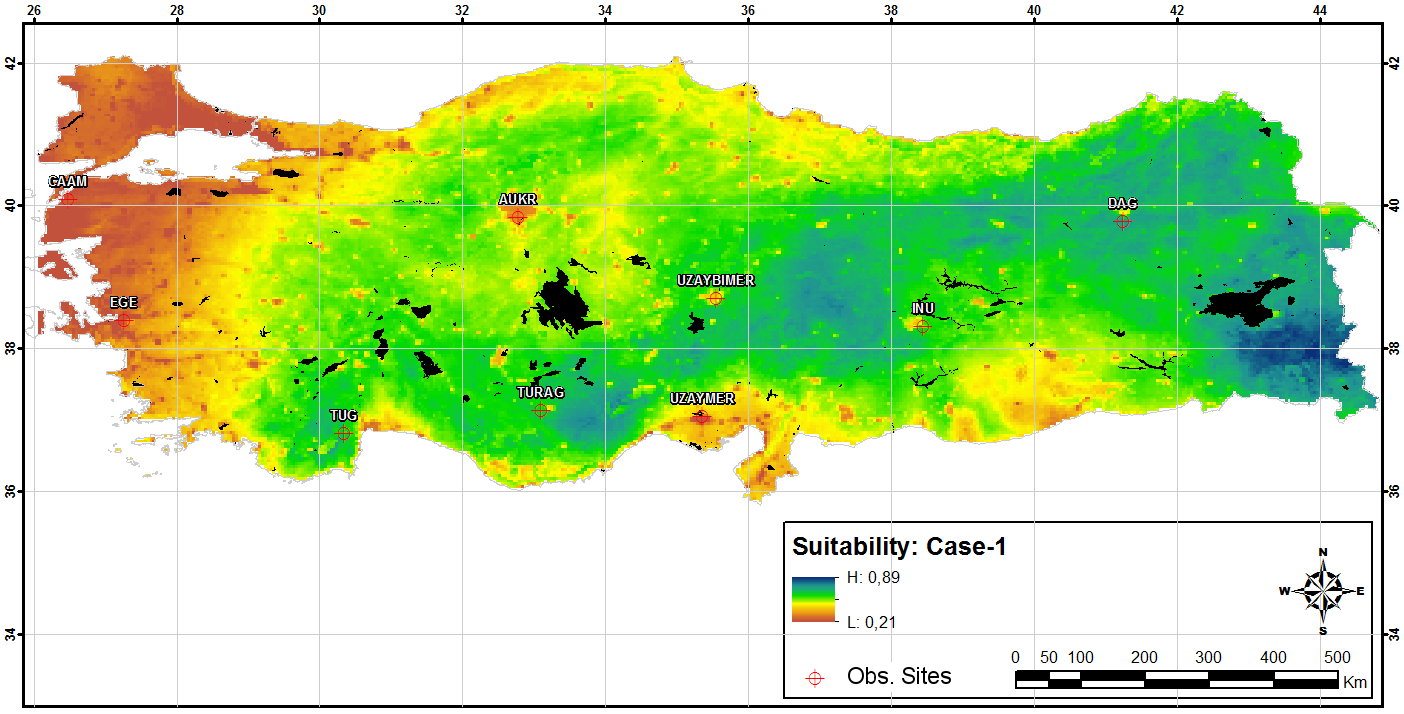}
\includegraphics[width=0.45\textwidth]{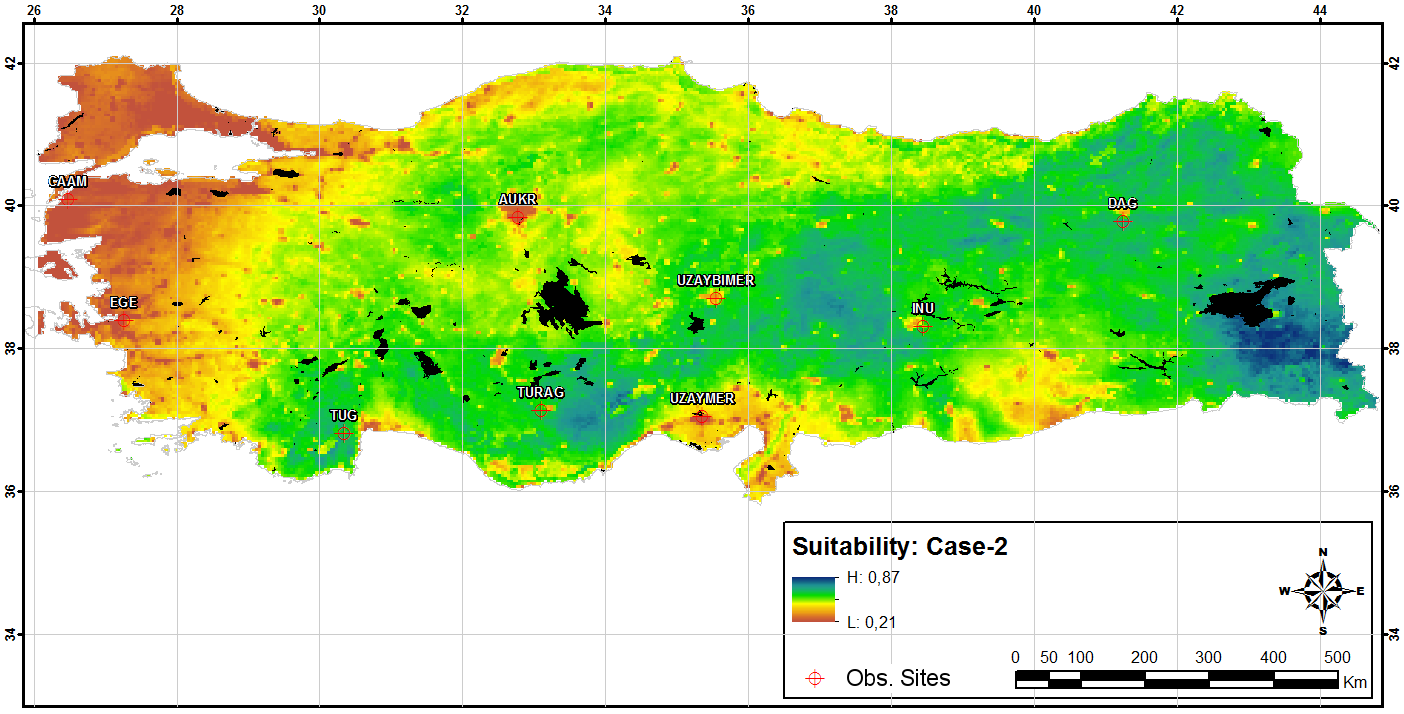}}
\cfig{%
\includegraphics[width=0.45\textwidth]{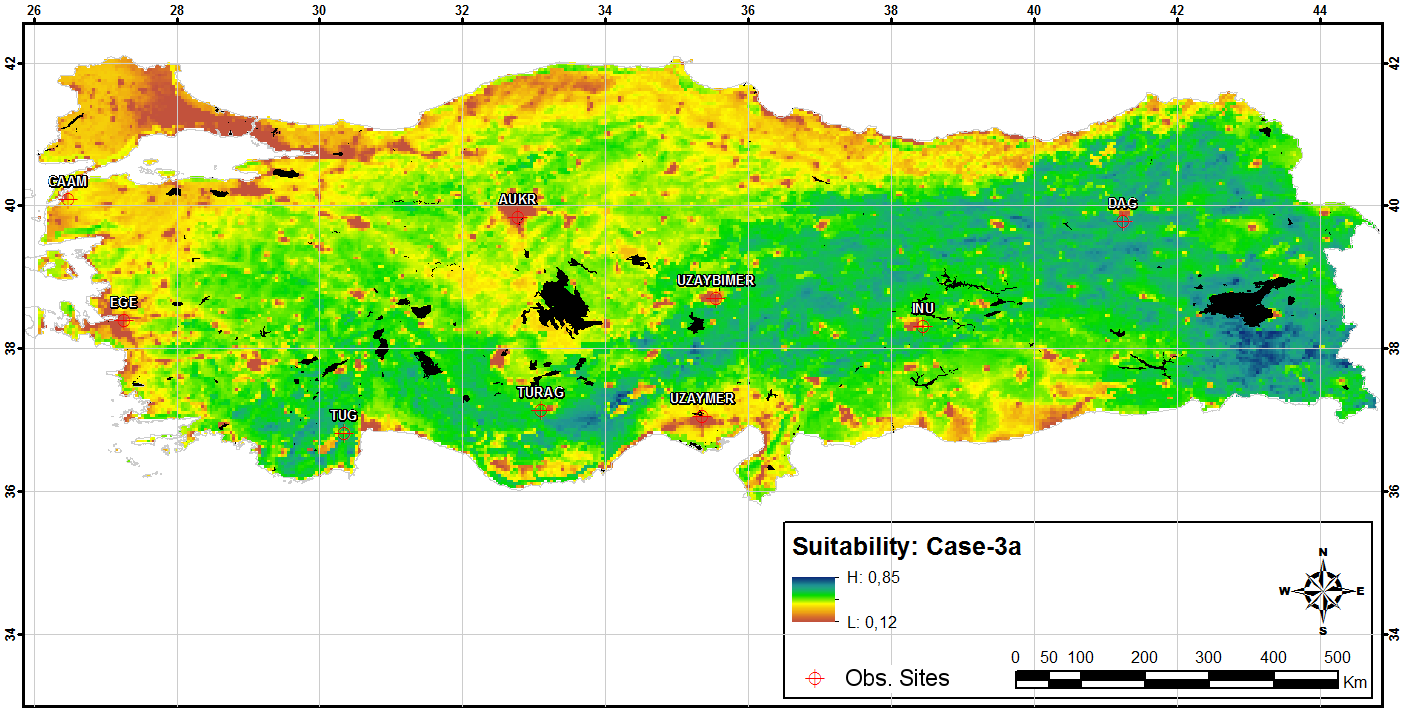}
\includegraphics[width=0.45\textwidth]{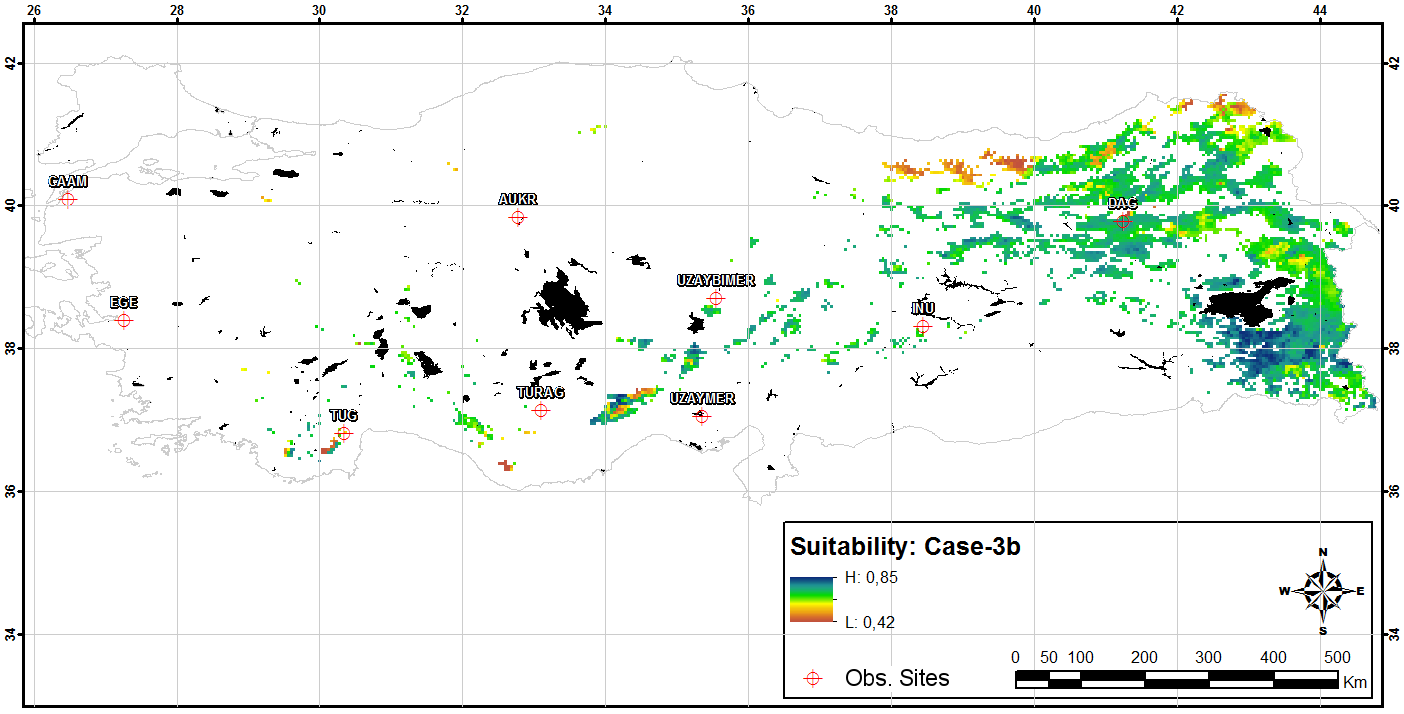}}
\cfig{%
\hspace*{0.45\textwidth}
\includegraphics[width=0.45\textwidth]{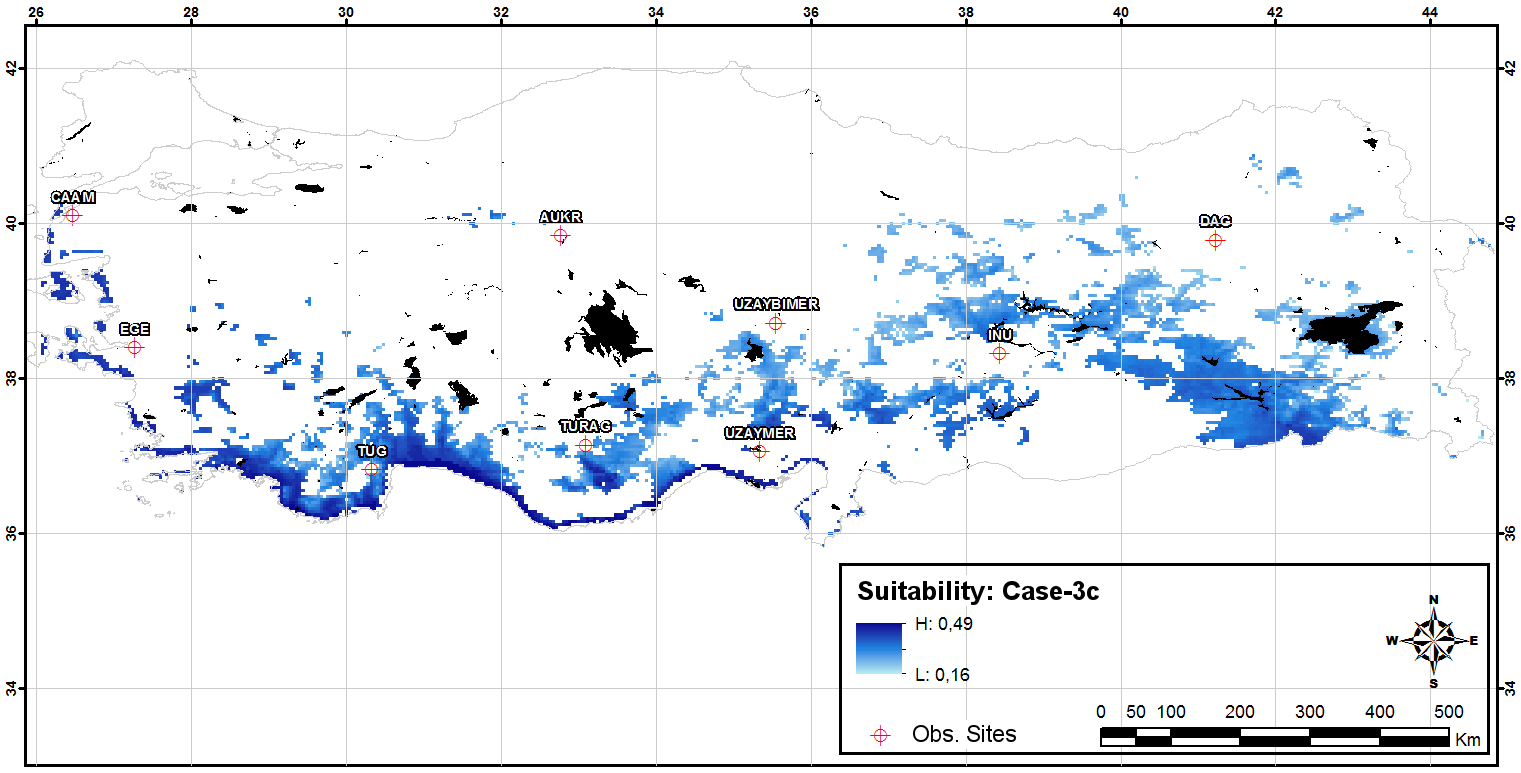}}
\caption{The results of MCDA for the cases listed in \S\ref{subsec:mcd}. Case-3 is divided into two. Case-3a: Improvement to Case-2; Case-3b: DEM values below 2000 m are filtered out; Case-3c: DEM values above 2000 m and CC values better than typical 8 weeks are filtered out.}
\label{f:casefig}
\end{figure*}

Further changes in the weights didn't change the results of analysis. Therefore Case-3 is taken as the final outcome of the analysis. The resultant cases (\ie maps) are given in Figure \ref{f:casefig}.
Even though it can be statistically scanned, a quick browse of the figure might give an almost complete impression of the good sites.

It has to be noted that even though site selection for an astronomical observatory is finalized with years long on-site tests which require dedicated funding, using GIS-MCDA analysis with relatively very small costs, a short list of potential sites could easily and accurately be created without physical contact with the sites themselves. MCDA is also a common method of relating very different layer informations among various data sets (\eg finding the best location of a shopping center in a city with varying pollution distributions). Therefore, in using MCDA, the only disadvantage of the method would be losing the relation between the data sets and \textit{physical} location itself, and suitability of the location is not confirmed until it is physically tested on-site. Other than this, the method has no disadvantages as a method itself.

\section{Results and Discussion}
\label{sec:results}

By browsing through all cases (see Figures \ref{f:casefig}) one can easily gather localization of potential regions throughout the study area (Table \ref{t:regions}).

To finalize the analysis some further discussions on the cases are needed. The histograms of both CC and DEM values (Figure \ref{f:hists}) will help in following the discussion:
\begin{figure}
\cfig{\includegraphics[width=0.45\textwidth]{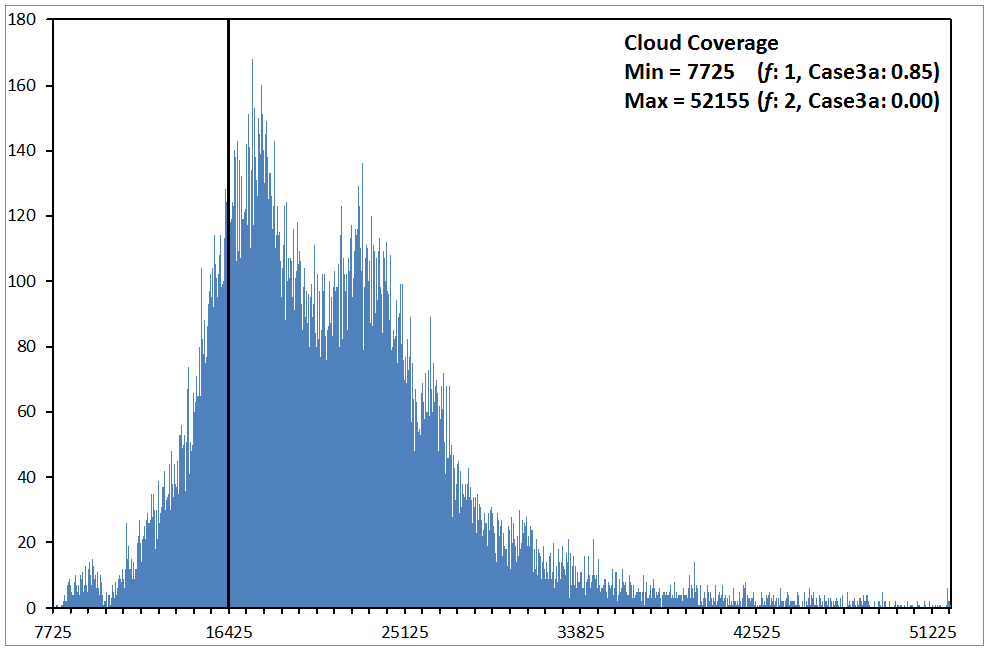}}
\cfig{\includegraphics[width=0.45\textwidth]{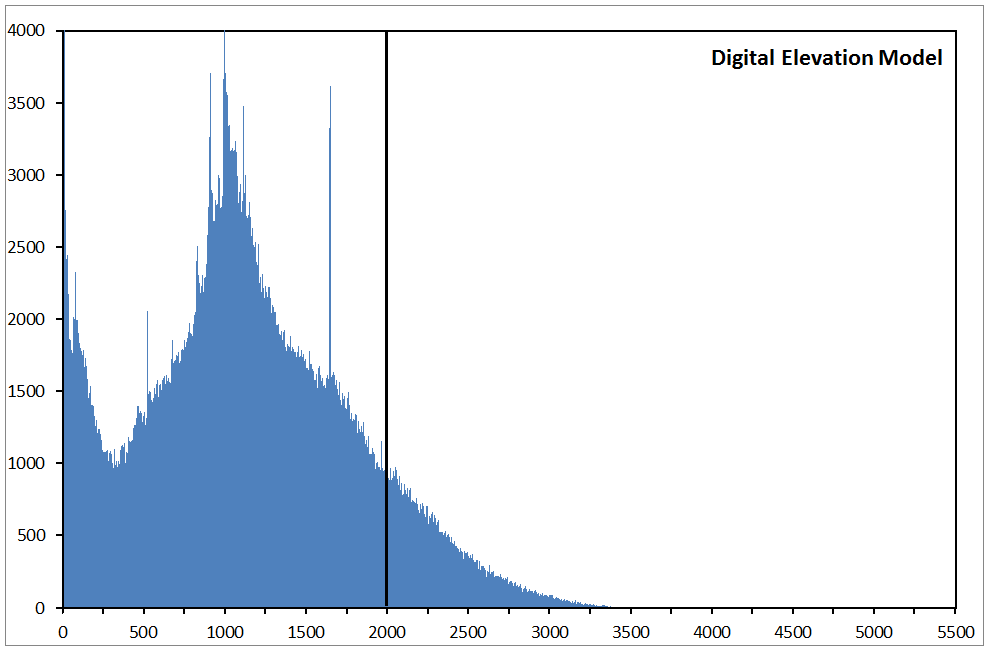}}
\caption{Histograms of both CC and DEM layers. Minimum, maximum, frequency and suitability value of the layer are given as a legend for CC; for DEM, these values are not applicable. A vertical line is drawn for CC representing minimum CC value of 8 weeks and for DEM representing the typical inversion layer of 2000 m.}
\label{f:hists}
\end{figure}

Even though it is a test case, new sites could successfully be located from the result of Case-1. Several of the site locations from Table \ref{t:siteinf} fall into this category (as well as the new ones).

For good sites, sky becomes transparent when the location has low CC, PWV and AOD values. As is explained above, the locations with high DEM values (which is strongly correlated with PWV) and having low CC (which is the fundamental driving factor of all layers) could be also taken as \textit{transparent sky} and therefore it leads us to the Case-3a.
In those selected sites satisfying these criteria, surveys requiring long-term studies and pointing observations requiring low astronomical seeing values could be carried out. Site locations of only TUG and DAG from Table \ref{t:siteinf} fall into this category.
Therefore, to spot those potentially low-seeing-value sites (which is strongly correlated with high DEM values), Case-3b is introduced by extracting out locations with DEM values $< 2000$ m leaving only high mountain ridges.

To confirm further that no potential site with \textit{low DEM values but with high CC suitability} condition are left out from Case-3b, Case-3c has been produced. In this case, we have started with only CC layer and \textit{subtracted} the effect of (a) DEM and AL (Case-3a); and (b) CC values above the minimum value of typical 8 weeks (see discussion of CC in \S\ref{subsec:ana} and Figure \ref{f:hists}). Note that due to the same reason explained in \S\ref{subsec:clo}, boundary between land surface and sea give false-positive results where they have to be ignored. Those sites falling to this special conditions are also suitable for long-term survey studies.
\section{Conclusion}
\label{sec:conc}

\begin{figure*}
\begin{tabular}{@{}c@{~~~~}c@{}}
\parbox[t][\textheight][t]{0.69\textwidth}{%
\includegraphics[angle=90,width=0.54\textheight]{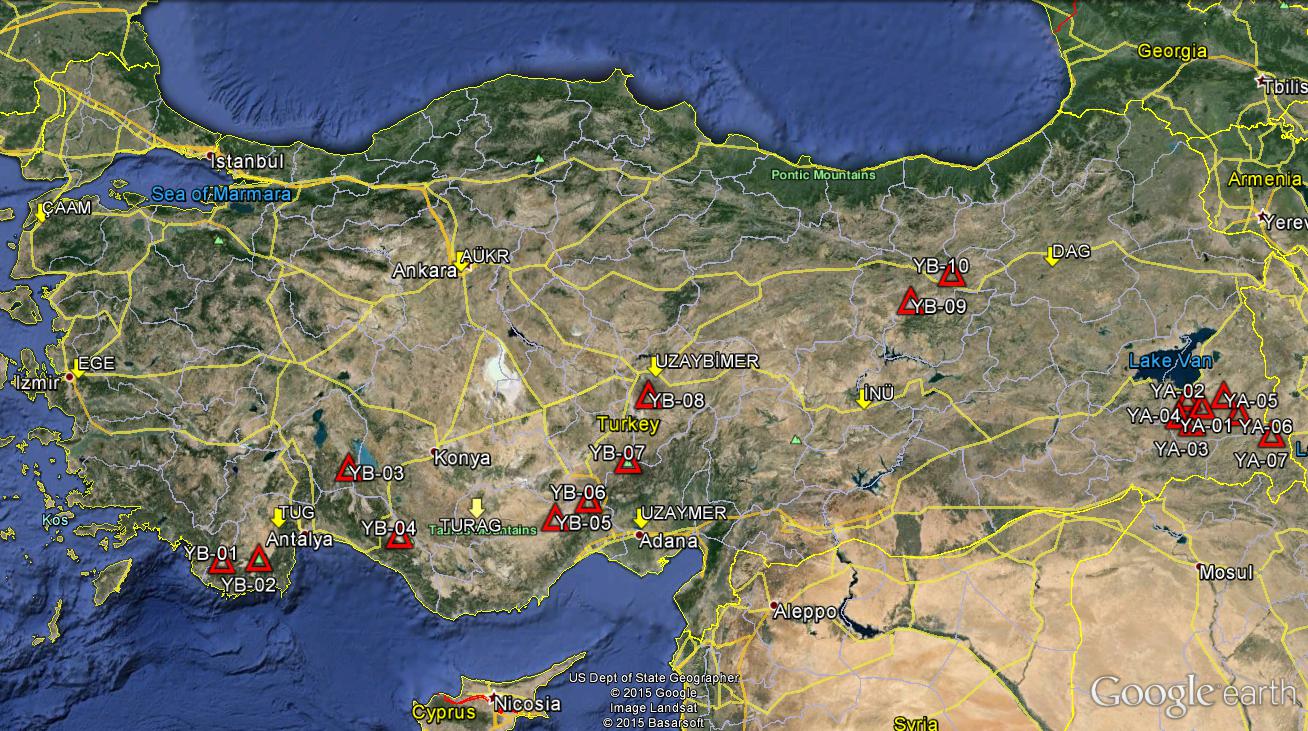}}
&\parbox[b][\textheight][b]{\textwidth}{%
\includegraphics[angle=90,width=0.21\textheight]{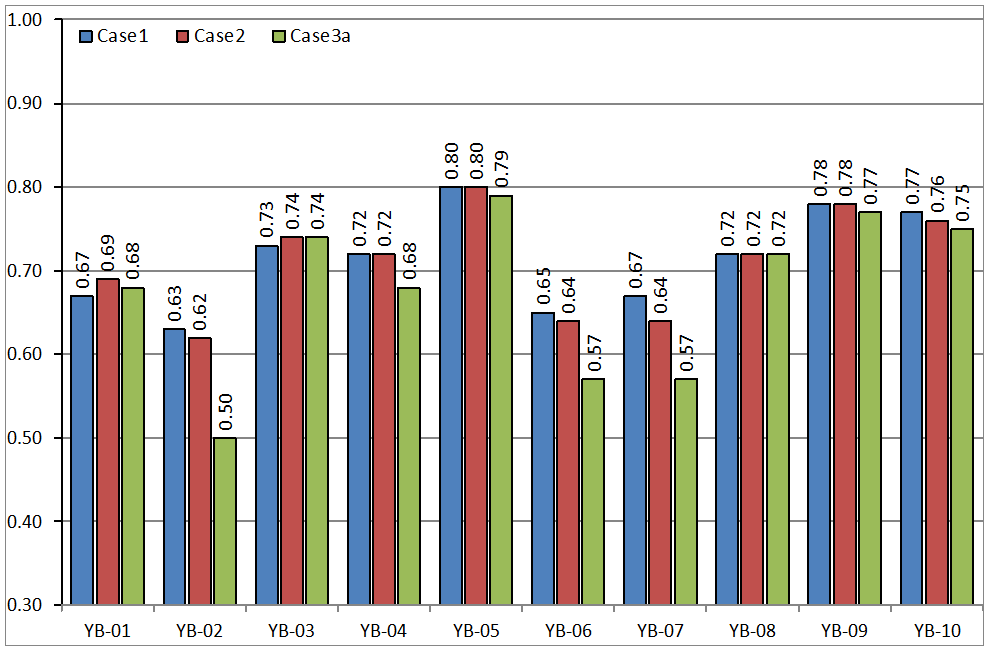}\vfil%
\includegraphics[angle=90,width=0.21\textheight]{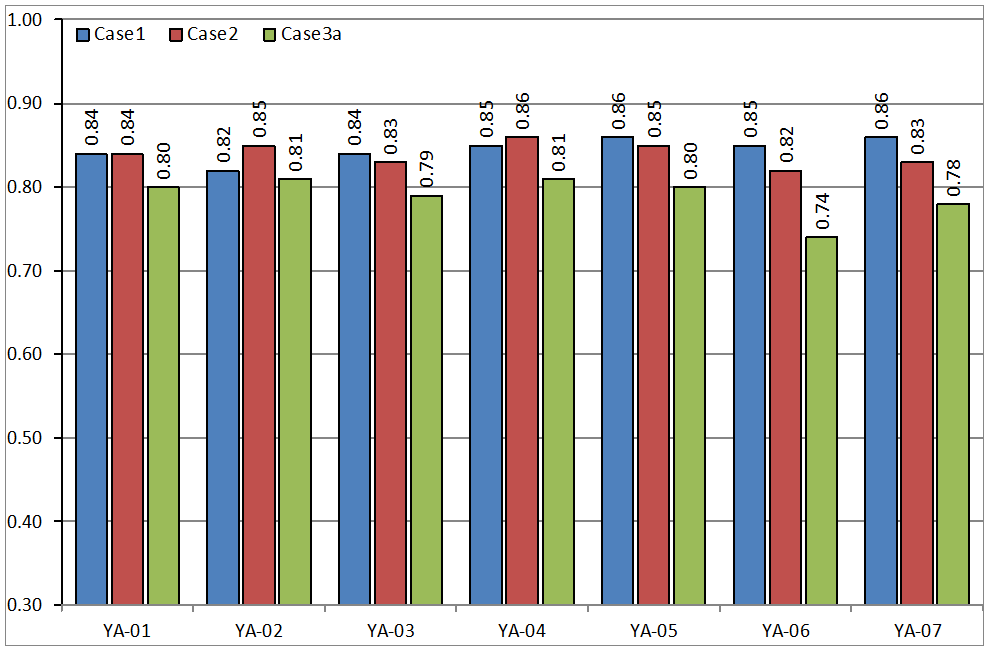}\vfil%
\includegraphics[angle=90,width=0.21\textheight]{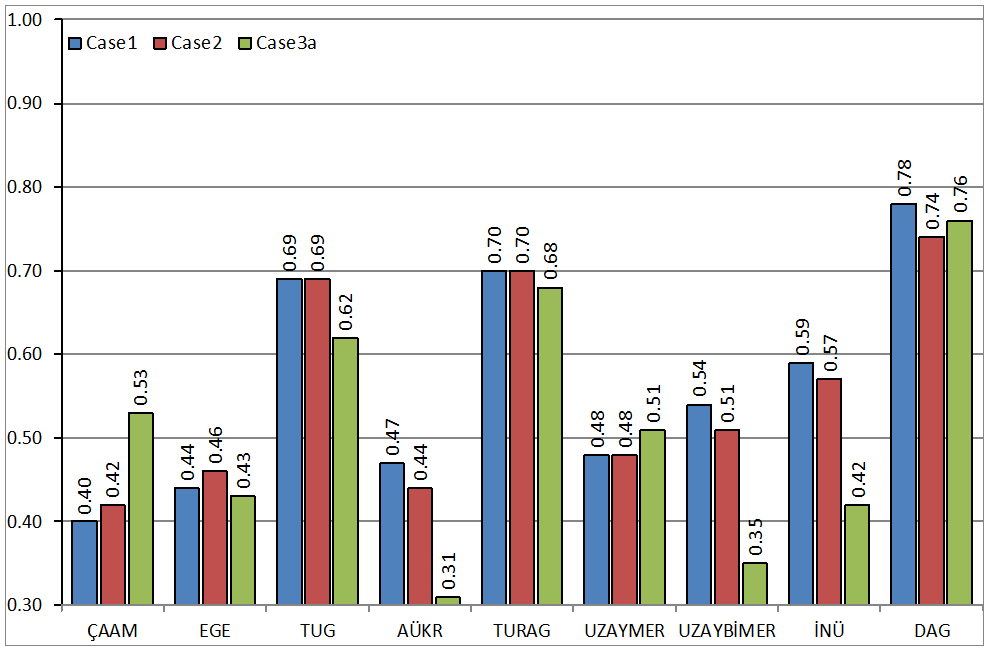}}
\end{tabular}
\caption{%
Map of sites (Left Panel).
17 suitable new observatory sites (red triangles) are shown in Turkish
geographic map.
The institutional observatory site locations (yellow arrows) are also shown.
The suitability values (Right Panel) of the site locations in Table
\ref{t:lists} are plotted for institutional sites (lowermost graph),
Region--A (middle graph) and Region--B (uppermost graph).
}
\label{f:casefigtr}
\end{figure*}

The results of MCDA can be summarized below:
\begin{itemize}
\item \textbf{Result-1:}  There is nowhere in Turkey with the suitability of value close to 1.0. The highest value is 0.85. Therefore, our major observatory sites such as TUG and DAG have quite good values. Furthermore, all new sites capable of hosting relatively large telescopes.

\item \textbf{Result-2:} These new sites can easily be divided into two major regions: Region-A (RA) containing YA-1 to YA-7 where all fall into southeast of Turkey; Region-B (RB) containing YB-1 to YB-10 where they can be followed though axial track of \textit{Taurus Mountains Ridge} from southwest to northeast of Turkey.
West of the axis along RB can easily be ignored due to their very low suitability values (\ie any site location from these regions will not be labeled as \textit{the best}).

\item All new positions, including the institutional observatory sites, are given in Table \ref{t:lists} and plotted in Figure \ref{f:casefigtr}. The criteria for this manual search of new locations were mainly due to logistic reasons (\eg having a pathway to the summit, close to civilization, reachable with a team of people) where decision can simply be made by eye.

\item \textbf{Result-3:} The RA is a remote\footnote{%
This outcome has been obtained from a commonly used demographic layer of Turkey. However, since the layer is not fully related to astronomy, it is not included in this study as an active layer.},
isolated region containing very high (around 3000-4000 m) mountain ridges with dry weather conditions. Their suitability are the highest for Turkey.

\item \textbf{Result-4:} Contrary to RA, there is no direct localization within RB positions.

\item \textbf{Result-5:} Case-3c reveals that (1) Case-3a is a good representation of new site locations; (2) the missed-good-locations are still close to the mountain ridges (but with lower DEM values) defined in Case-3a (\ie RA and RB).

\item \textbf{Result-6:} The real \textit{best} sites can only be found after the locations have been throughly tested on-site and/or local MCDA-GIS studies.

\item Since our short list is relatively long to come with a final result, on-site tests should be done at once for several locations. Therefore, we have applied for a national research funding (and awarded since March 2014) to produce a prototype of an on-site test unit (to measure both astronomical and meteorological parameters) which might be used on these locations. The usage and production of the unit has to be further realized elsewhere.

\item Even though their wavelength and their target were different, we confirm the results of \cite{2012ExA....33....1K} that TURAG site (around YB-05 and YB-06) is well above the suitability of others for an optical observatory site.
\end{itemize}
\begin{table}
\centering
\caption{List of institutional observatory sites (upper panel) and new suitable observatory sites (lower two panels) in Turkey. The new sites are divided into regions which are explained in section \S\ref{sec:conc}.}
\label{t:lists}
\begin{tabular}{@{}l@{~}c@{~}c@{~}c@{~}c@{~}c@{~}c@{}}
\hline
\noalign{\smallskip}
\multicolumn{1}{c}{Name}
	& Longitude
	& Latitude
	& Elevation
	& Case1
	& Case2
	& Case3\\
\multicolumn{1}{c}{(Region)}
	& (\dgr~ East)
	& (\dgr~ North)
	& (m)
	&
	&
	& \\ \hline
\noalign{\smallskip}
\c{C}AAM    & 26.48 & 40.10 &  373 & 0.40 & 0.42 & 0.53\\
EGE         & 27.27 & 38.40 &  622 & 0.44 & 0.46 & 0.43\\
TUG         & 30.34 & 36.82 & 2436 & 0.69 & 0.69 & 0.62\\
A\"UKR      & 32.78 & 39.84 & 1254 & 0.47 & 0.44 & 0.31\\
TURAG       & 33.09 & 37.14 & 1062 & 0.70 & 0.70 & 0.68\\
UZAYMER     & 35.35 & 37.06 &  112 & 0.48 & 0.48 & 0.51\\
UZAYB\.IMER & 35.55 & 38.71 & 1094 & 0.54 & 0.51 & 0.35\\
\.IN\"U     & 38.44 & 38.32 & 1021 & 0.59 & 0.57 & 0.42\\
DAG         & 41.23 & 39.78 & 3102 & 0.78 & 0.74 & 0.76\\
\noalign{\smallskip}
\hline
\noalign{\smallskip}
YA-01       & 42.95 & 38.06 & 3493 & 0.84 & 0.84 & 0.80\\
YA-02       & 43.11 & 38.24 & 3515 & 0.82 & 0.85 & 0.81\\
YA-03       & 43.14 & 37.97 & 3301 & 0.84 & 0.83 & 0.79\\
YA-04       & 43.31 & 38.14 & 2847 & 0.85 & 0.86 & 0.81\\
YA-05       & 43.65 & 38.23 & 3380 & 0.86 & 0.85 & 0.80\\
YA-06       & 43.85 & 38.03 & 3576 & 0.85 & 0.82 & 0.74\\
YA-07       & 44.32 & 37.75 & 3531 & 0.86 & 0.83 & 0.78\\
\noalign{\smallskip}
\hline
\noalign{\smallskip}
YB-01       & 29.57 & 36.54 & 2939 & 0.67 & 0.69 & 0.68\\
YB-02       & 30.08 & 36.59 & 2230 & 0.63 & 0.62 & 0.50\\
YB-03       & 31.30 & 37.66 & 2866 & 0.73 & 0.74 & 0.74\\
YB-04       & 32.03 & 36.94 & 2471 & 0.72 & 0.72 & 0.68\\
YB-05       & 34.18 & 37.19 & 2754 & 0.80 & 0.80 & 0.79\\
YB-06       & 34.63 & 37.39 & 3499 & 0.65 & 0.64 & 0.57\\
YB-07       & 35.17 & 37.83 & 3504 & 0.67 & 0.64 & 0.57\\
YB-08       & 35.45 & 38.53 & 3835 & 0.72 & 0.72 & 0.72\\
YB-09       & 39.16 & 39.50 & 3316 & 0.78 & 0.78 & 0.77\\
YB-10       & 39.77 & 39.79 & 3426 & 0.77 & 0.76 & 0.75\\
\hline
\end{tabular}
\end{table}
\begin{table}
\centering
\caption{Regional distribution analysis of the new sites from Case3a.
To validate the results of manual search of new sites, a circular region with a radius of 40 km (equivalent to $\simeq$10 pixels with the layer resolution) has been chosen at that geographic location.
Statistics of suitability distribution within this circle is given as minimum, maximum and difference of these two values (2nd, 3rd and 4th columns, respectively).
The last column should be as low as possible for a good site where suitability remains unchanged for a very wide surface area.
These values are represented in the right-most bar chart of Figure \ref{f:casefigtr}.
}
\label{t:regions}
\begin{tabular}{@{}clccc@{}}
Region
	& \multicolumn{3}{c}{Suitability}\\
        & Min & Max & $\Delta$ \\
\hline
\noalign{\smallskip}
YA-01 & 0.60 & 0.82 & 0.22\\
YA-02 & 0.44 & 0.84 & 0.40\\
YA-03 & 0.65 & 0.84 & 0.19\\
YA-04 & 0.63 & 0.82 & 0.19\\
YA-05 & 0.41 & 0.81 & 0.40\\
YA-06 & 0.73 & 0.81 & 0.04\\
YA-07 & 0.64 & 0.77 & 0.13\\
\hline
\noalign{\smallskip}
YB-01 & 0.37 & 0.78 & 0.41\\
YB-02 & 0.37 & 0.79 & 0.42\\
YB-03 & 0.57 & 0.80 & 0.23\\
YB-04 & 0.39 & 0.83 & 0.44\\
YB-05 & 0.38 & 0.82 & 0.44\\
YB-06 & 0.55 & 0.76 & 0.21\\
YB-07 & 0.42 & 0.77 & 0.35\\
YB-08 & 0.34 & 0.77 & 0.43\\
YB-09 & 0.54 & 0.75 & 0.21\\
YB-10 & 0.34 & 0.75 & 0.41\\
\hline
\noalign{\smallskip}
\end{tabular}
\end{table}
\begin{acknowledgements}
This research is funded through the TUBITAK project MFAG-113F266.
The authors thank to the Turkish Astronomical Society for the support in establishing the national synergie for the project.
Each author thanks to their home university for their support in this national project.
We are grateful to the reviewer for their constructive comments and recommendations.
The authors thank to: TUG (TUBITAK National Observatory) for its support to the project and usage of its facilities; DAG (Dogu Anadolu Gozlemevi - Eastern Anatolia Observatory) and Atat\"urk University (through DPT Project No: 2011K120230) for their support in initiating a general GIS solution to the problem and support given to project; Turkish State Meteorological Service for the data sets that they have provided.
The authors acknowledge the MODIS Science team for the Science Algorithms, the Processing Team for producing MODIS data, and the GES DAAC MODIS Data Support Team for making MODIS data available to the user community.
Image and data processing by NOAA's National Geophysical Data Center.
DMSP data collected by US Air Force Weather Agency.
ASTER GDEM is a product of METI and NASA.
%
\end{acknowledgements}


  \label{lastpage}
\end{document}